# The Contribution of Active Galactic Nuclei to the Diffuse Gamma Ray Background

P.A. Johnson[1], A. Mastichiadis[2], R.J. Protheroe[1], T.S. Stanev[3] and

A.P. Szabo[1]*

[1]Department of Physics and Mathematical Physics, University of Adelaide,

SA 5005, Australia

[2]Max-Planck-Institut für Kernphysik, Heidelberg, Germany.

[3]Bartol Research Institute, University of Delaware, Newark, DE 19716,

U.S.A.

*Now at Defence Science and Technology Organization, P.O. Box 1500,

Salisbury 5108, Australia.

Short Title: Gamma Rays from Active Galactic Nuclei

Classification Numbers: 9870R, 9870V, 9850

**Abstract**

We consider the emission of high energy to very high energy $\gamma$-rays in radio-quiet active galactic nuclei (AGN) or the central regions of radio-loud AGN. We use our results to estimate the $\gamma$-ray flux from the central regions of nearby AGN, and then to calculate the contribution to the diffuse $\gamma$-ray flux from unresolved AGN.

# 1  Introduction

The Compton Gamma Ray Observatory has currently detected at least 21 AGN at GeV energies (Fichtel 1994), and one of these, Mrk 421, has now also been observed at TeV energies (Punch *et al.* 1992). All the objects detected so far are blazars, and the $\gamma$-ray emission is almost certainly from the jet. Models for $\gamma$-ray emission from AGN jets have recently been proposed, and involve inverse-Compton scattering of synchrotron emission or external photons, possibly from an accretion disk (Dermer, Schlickeiser and Mastichiadis 1992; Protheroe, Mastichiadis and Dermer 1992; Sikora, Begelman and Rees 1994; Zdziarski and Krolik 1993), or pair-synchrotron cascades resulting from proton acceleration (Mannheim 1993a,b). Blazars, however, form a small fraction of all AGN, and we wish to consider the $\gamma$-ray emission from the much more numerous 'ordinary' AGN (e.g. Seyferts and radio-quiet AGN), or from the central engines of blazars as distinct from the jets.

Because of the high radiation density near the central engines of AGN most $\gamma$-rays from this region suffer photon-photon pair production interactions. However, production of high energy neutrons in the central region provides a way of transporting energy out to radii from which $\gamma$-rays can escape (Kirk and Mastichiadis 1989, Sikora *et al.* 1989, Mastichiadis and Protheroe 1990, Giovanoni and Kazanas 1990, Atoyan 1992a,b). Neutrons can be produced in $p\gamma$ collisions in which photoproduction takes place, and in $pp$ collisions. In the case of $p\gamma$ interactions, most interactions take place near threshold where the $\Delta$ resonance dominates. Mastichiadis and Protheroe (1990) have used the neutron production spectrum estimated by Kirk and Mastichiadis (1989), along with an approximate semi-analytic treatment, to calculate the $\gamma$-ray flux from AGN. Kirk and Mastichiadis (1989) made some simplifying assumptions in order to derive their results in an analytic way. For example, they only considered interactions with radiation, assumed an average value of the pion photoproduction cross section, and related the energy of the interacting proton with that of the produced neutron by a delta-function. Here we use the neutron production spectrum from a more rigorous Monte Carlo treatment of acceleration and particle interactions in the central regions of AGN (Szabo and Protheroe 1992a;1994a,b) which includes interactions with both matter and radiation as well as a treatment of diffusive escape of accelerated particles and advection into the black hole. Furthermore, in the model used in the present work the proton acceleration rate and the rate of injection of protons into the accelerator, together with the matter, radiation and magnetic environments, are related to the AGN continuum luminosity in a consistent way.

To calculate the high energy $\gamma$-ray emission, we consider the cascade initiated by high energy electrons and photons produced in high energy interactions in the central region and farther out to calculate the $\gamma$-ray emission from AGN. A brief description of this work together with preliminary results is given by Johnson *et al.* (1993a,b).

## 2   The model

The AGN model we adopt is that described by Protheroe and Kazanas (1983) and developed by Kazanas and Ellison (1986). The basic ingredients of the model are summarized below. A shock at radius $R = x_1 r_S$, where $r_S$ is the Schwarzschild radius, is assumed to develop in an accretion flow onto a supermassive black hole and be supported by the pressure of relativistic particles. We therefore assume that at the shock the magnetic pressure is comparable to that of relativistic particles and the ram pressure of accreting plasma:

$$\rho u_1^2 \simeq U_p/3 \simeq B^2/8\pi, \quad (1)$$

where $U_p$ is the energy density in relativistic particles and $u_1 = x_1^{-1/2} c$ is the upstream flow velocity. We assume that relativistic particles are responsible for the AGN continuum luminosity, $L_C$, and that about half their energy is lost to neutrinos, i.e.

$$L_C \simeq \frac{1}{2} L_p. \quad (2)$$

The luminosity in relativistic particles is given by

$$L_p \simeq \epsilon \frac{1}{2} \dot{M} u_1^2 \quad (3)$$

where $\epsilon(x_1) = 1 - 0.1 x_1^{0.31}$ is the efficiency of conversion of bulk kinetic energy of accreting plasma into energetic particles at the shock (Kazanas and Ellison 1986). In this model, the black hole mass is proportional to luminosity, and from Figure 5 of Kazanas and Ellison (1986) we obtain

$$\dot{M} \simeq 10^{-38} x_1 L_C \quad M_\odot \qquad (4)$$

where $L_C$ is in erg s$^{-1}$.

We are now in a position to specify the matter, magnetic and radiation environment in which the particles are accelerated and interact. The matter density at the shock is related to the accretion rate by

$$\dot{M} = 4\pi R^2 \rho u_1. \qquad (5)$$

Hence, we obtain the matter density

$$\rho \simeq 1.4 \times 10^{33} \epsilon^{-1} x_1^{-5/2} L_C^{-1} \quad \text{g cm}^{-3}, \qquad (6)$$

and the magnetic field

$$B \simeq 5.5 \times 10^{27} \epsilon^{-1/2} x_1^{-7/4} L_C^{-1/2} \quad \text{gauss}. \qquad (7)$$

Later, when dealing with synchrotron radiation by electrons we will need also the magnetic field at large radii. For this, we follow Protheroe and Szabo (1992) and assume that the magnetic pressure tracks the plasma pressure giving $B \propto r^{-5/4}$. The radiation density in the vicinity of the shock is related to the AGN continuum luminosity by

$$U_{\text{rad}} \simeq L_C / \pi R^2 c \qquad (8)$$

giving

$$U_{\rm rad} \simeq 1.2 \times 10^{54} x_1^{-4} L_C^{-1} \quad {\rm erg \ cm}^{-3}. \qquad (9)$$

Making the approximation that the diffusion coefficient at the shock is some constant $b$ times the Bohm diffusion coefficient, i.e. $D = b(\frac{1}{3} r_g c)$ where $r_g$ is the gyroradius, we find that

$$\frac{{\rm d}E_p}{{\rm d}t} \simeq 2.5 \times 10^{39} b^{-1} \epsilon^{-1/2} x_1^{-11/4} L_C^{-1/2} \quad {\rm eV \ s}^{-1}. \qquad (10)$$

The main loss processes for high energy protons in the central regions of AGN are $p\gamma$ collisions with photons of the infrared–X-ray AGN continuum resulting in either $e^{\pm}$ pair production or pion production, $pp$ collisions with protons of the accreting plasma, diffusive escape from the acceleration region, advection into the black hole, and synchrotron losses. Szabo and Protheroe (1991, 1992b, 1994a,b) describe these processes which occur during and after acceleration of protons in a radiation field (see also Sikora *et al.* 1989, and Sikora and Begelman 1992). Szabo and Protheroe (1992a) have applied this to acceleration in the central regions of AGN to calculate the diffuse high energy neutrino intensity from unresolved AGN (see also Stecker *et al.* 1991), and have also found that high energy neutron production could give rise to an observable component of the cosmic ray intensity at $10^7$ GeV (Protheroe and Szabo 1992).

The maximum proton energy $E_{\mathrm{max}}$, occurs where the acceleration rate equals the total energy-loss rate and is given by Szabo and Protheroe (1994a,b). As discussed by Szabo and Protheroe (1992a), we consider two classes of continuum spectra, which bracket the observed spectra, referred to here as A and B. Both contain a UV thermal bump, but B has a flat equal–energy–per–decade photon spectrum (power law index –2.0) extending from $\sim$ 1 MeV down into the infrared, whilst A is infrared deficient, with a harder power law of index –1.7 extending only above the UV bump up to $\sim$ 1 MeV. The thermal bump in spectrum A has a temperature of $5 \times 10^4$K, whilst that of B is slightly cooler at $3 \times 10^4$K. In both cases, equal energy density is assumed to be contained in the power-law and the black-body components. $E_{\mathrm{max}}$ is given for both assumed spectra (A and B) in Figure 1, where it is plotted against the AGN continuum luminosity, $L_C$, divided by $b^2$.

# 3 Production spectra of electrons and photons

In the simulations, which are described fully by Szabo and Protheroe (1992a, 1992b, 1994a,b), the maximum proton energy is adopted as a parameter, and the acceleration rate is set to equal the loss rate at $E_{\rm max}$. We may visualize the shock accelerator as a leaky box in which particles of energy $E_p$ are continually accelerated at a constant rate $a$, independent of energy, and have a probability of escaping from the box per unit time of $a/E_p$.

In the computer code, we inject protons into the 'leaky box' and simulate their acceleration, interactions and escape from the accelerator. We also deal with the interactions of protons which remain magnetically confined in the central region after leaving the acceleration region. In $p\gamma$ and $pp$ collisions, there is some probability of producing neutrons in the final state. These neutrons, not being magnetically confined provide a mechanism for transporting energy away from the intense radiation field in the central region because the straight-line optical depth to $np$ collisions in the accreting plasma is much less than one (Szabo and Protheroe 1992a). Some fraction of protons from neutron decay may escape from the host galaxy and contribute to the pool of extragalactic cosmic rays (Protheroe and Szabo 1992). However, below $10^5$ GeV the bulk of the neutrons decay in a region of enhanced matter density in the accretion flow and the resulting protons are likely to suffer multiple $pp$ collisions giving rise to pion production. We model the $\pi \to \mu \to e$ and $\pi^0 \to \gamma\gamma$ decays to obtain the spectra of $\gamma$-rays and electrons produced as a function of radius as a result of injecting one proton into the accelerator. Examples of these production spectra are given in Figure 2. Since $E_{max}$ is defined as the energy at which the acceleration rate equals the total energy-loss rate, it is of course possible for protons to achieve higher energies than this value. This is illustrated in Figure 2 where the electron and $\gamma$−ray spectrum continues past the nominal maximum energy of $10^8$ GeV.

## 4   The pair–synchrotron cascade

The importance of synchrotron radiation by electrons from photon-photon pair production and $\pi \to \mu \to e$ decay for VHE $\gamma$-ray emission in AGN was pointed out by Mastichiadis and Protheroe (1990). This is because in the energy region in which cascading is important, i.e. where photon–photon pair production takes place, inverse-Compton scattering is in the Klein-Nishina regime, and for the comparable energy densities in the radiation field and magnetic field which apply in the present work, synchrotron energy losses dominate over inverse-Compton losses. In the present work we assume that all $\gamma$-rays emerging from the AGN result from a pair–synchrotron cascade fed by electrons and $\gamma$-rays produced as a result of in $p\gamma$ and $pp$ collisions in the central region, and $pp$ collisions of protons from neutron decay. The photon-field of the AGN can be characterized by a compactness parameter, useful in facilitating the calculation of interaction lengths, defined by

$$l = \frac{L_C \sigma_T}{4\pi R m_e c^3} \qquad (11)$$

In the present model, $l$ can be simply expressed in terms of $x_1$ only,

$$l = \frac{730}{x_1^2} \qquad (12)$$

as a consequence of Equation (4) (Szabo and Protheroe 1992a). The available data (Wandel and Yahil, 1985), after applying a bolometric correction of $\times 5$, suggest that $x_1$ values in the range 10 to 100, and a flat distribution in $\log x_1$ for $10 < x_1 < 100$ appears to be consistent with the data. When calculating the various optical depths, we do so for unit compactness parameter, and then those for other compactness parameters are easily found by scaling.

To solve the cascade equations we consider the contribution to the emerging $\gamma$-ray luminosity from the spherical shell of radius $r$ to $(r + \mathrm{d}r)$. Let the production spectra of electrons (initially from $\pi \to \mu \to e$ decay) and photons (initially from $\pi^0 \to \gamma\gamma$ decay) of energy $E$ in this shell per injected proton be $q_e(E, r)\mathrm{d}r$ and $q_\gamma(E, r)\mathrm{d}r$ respectively. The production spectra in Figure 2 show

$$\frac{\mathrm{d}n}{\mathrm{d}E} = \int_{r_{\mathrm{in}}}^{r_{\mathrm{out}}} \mathrm{d}r q(E, r) \tag{13}$$

where $q(E, r)$ is either $q_e(E, r)$ or $q_\gamma(E, r)$. We will consider separately the pair–synchrotron cascade in the central region and outside the central region.

## 4.1 Outside the central region

Outside the central region we treat the pair–synchrotron cascade in the following way. First, we use functions described by Protheroe (1990) to obtain the synchrotron yield of photons of energy $E$ per electron of energy $E'$ injected into the magnetic field appropriate to radius $r$, $Y_{\mathrm{syn}}(E, E', r)$. Hence we add to $q_\gamma(E, r)\mathrm{d}r$ the production spectrum of photons from synchrotron radiation by electrons produced in the spherical shell,

$$q_\gamma(E, r)\mathrm{d}r \leftarrow q_\gamma(E, r)\mathrm{d}r + \int_E^\infty \mathrm{d}E' [Y_{\mathrm{syn}}(E, E', r) q_e(E', r)\mathrm{d}r], \tag{14}$$

where the left-pointing arrow is used to indicate that the new iteration of the quantity on the left is given by the present value of the quantity on the right.

We then calculate the optical depth of $\gamma$-rays produced at radius $r$ against photon-photon pair production as a function of direction of emission (angle relative to the radially outward direction) in the infrared to hard X-ray AGN continuum as described by Mastichiadis and Protheroe (1990). This radiation field is anisotropic outside of the central region, and the optical depths are much less for photons emitted outwards compared with those emitted towards the centre (see also Protheroe *et al.* 1992). We assume that the $\gamma$-rays are emitted isotropically because an interacting proton's direction will have been randomized by the magnetic field, and we use the optical depths to calculate the average escape probability of $\gamma$-rays of energy $E$ emitted at radius $r$,

$$P_{\rm esc}(E,r) = \frac{1}{4\pi} \int d\Omega \exp[-\tau_{\gamma\gamma}(E,r,\theta)], \qquad (15)$$

where $\tau_{\gamma\gamma}(E,r,\theta)$ is the optical depth for a photon of energy $E$, produced at radius $r$ travelling at angle $\theta$ to the radially outward direction. The resulting escape probability $P_{\rm esc}(E,r)$, shown in Figure 3 for Spectrum A.

We then obtain our initial estimate (neglecting cascading) of the spectrum of escaping $\gamma$-rays per injected proton,

$$q_\gamma^{\rm esc}(E,r){\rm d}r = P_{\rm esc}(E,r)q_\gamma(E,r){\rm d}r. \qquad (16)$$

To obtain an improved estimate, we must take account of synchrotron radiation by the electrons from photon-photon pair production. In photon-photon pair production, one electron always carries most of the energetic photon's energy, and hence we make the approximation that the effect of pair production by a photon of energy $E$ is simply to convert it to an electron of energy $E$. The photons which pair produce then give a new electron production spectrum,

$$q_e(E,r){\rm d}r \leftarrow [1 - P_{\rm esc}(E,r)]q_\gamma(E,r){\rm d}r, \qquad (17)$$

and hence a new production spectrum of photons from synchrotron radiation from the spherical shell,

$$q_\gamma(E,r){\rm d}r \leftarrow \int_E^\infty {\rm d}E'[Y_{\rm syn}(E,E',r)q_e(E',r){\rm d}r]. \qquad (18)$$

Thus, the second approximation to the escaping $\gamma$-ray spectrum per injected proton becomes

$$q_\gamma^{\rm esc}(E,r){\rm d}r \leftarrow q_\gamma^{\rm esc}(E,r){\rm d}r + P_{\rm esc}(E,r)q_\gamma(E,r){\rm d}r. \qquad (19)$$

To obtain better estimates, Equations 17, 18 and 19 are iterated until the relative change in the calculated spectra is negligible. Finally, integration over radius yields the escaping $\gamma$-ray spectrum per injected proton due to interactions outside the central region,

$$Q_\gamma^{\text{out}}(E) = \int_R^\infty \mathrm{d}r\, q_\gamma^{\text{esc}}(E,r). \tag{20}$$

The contribution to the total escaping $\gamma$-ray spectrum from various shells,

$$\frac{\mathrm{d}n}{\mathrm{d}E} = \int_{r_{\text{in}}}^{r_{\text{out}}} \mathrm{d}r\, q_\gamma^{\text{esc}}(E,r), \tag{21}$$

is given in Figure 4 for the case where $E_{\text{max}} = 10^8\,\text{GeV}$. Also shown (dashed) is the result that would be obtained if cascading were ignored,

$$\frac{\mathrm{d}n}{\mathrm{d}E} = \int_{r_{\text{in}}}^{r_{\text{out}}} \mathrm{d}r\, q_\gamma^\pi(E,r) P_{\text{esc}}(E,r), \tag{22}$$

where $q_\gamma^\pi(E,r)$ is the initial spectrum of $\gamma$-rays resulting from $\pi^0$ decay.

## 4.2 Inside the central region

We assume that inside the central region of radius $R$ the radiation field is uniform and isotropic with energy density given by Equation 8. We then calculate the optical depth due to pair–production for traversing a distance $R$ in such a field, $\tau_{\gamma\gamma}^{\text{in}}(E)$. By solving the equation of radiative transfer we find the average probability of $\gamma$-rays produced uniformly throughout the central region escaping into the outer region without interaction,

$$P_{\text{esc}}^{\text{in}}(E) = \frac{3}{4\tau_{\gamma\gamma}^{\text{in}}(E)} \left[1 - \exp\left(\frac{-4\tau_{\gamma\gamma}^{\text{in}}(E)}{3}\right)\right]. \tag{23}$$

Multiplying by the probability of escaping from radius $R$ to infinity gives the overall average escape probability for photons produced inside the central region,

$$P_{\text{esc}}(E, \leq R) = 2P_{\text{esc}}^{\text{in}}(E)P_{\text{esc}}(E, R), \tag{24}$$

where for $r = R$, the escape probability for a $\gamma$-ray emitted in the outward hemisphere is approximately twice the average probability for escape given by equation 15, giving rise to the extra factor of 2. We can now calculate the spectrum of $\gamma$-rays emerging from the central region without interaction.

Throughout the central region, we take the magnetic field to be equal to that at $r = R$, and making the same approximation as before for the spectrum of electron-positron pairs produced, we calculate the additional contribution from synchrotron radiation, and iterate as before to obtain the escaping $\gamma$-ray spectrum per injected proton due to interactions inside the central region, $Q_\gamma^{\text{in}}(E)$, and hence the total spectrum per injected proton,

$$Q_\gamma(E) = Q_\gamma^{\text{in}}(E) + Q_\gamma^{\text{out}}(E). \tag{25}$$

The contribution to the total escaping $\gamma$-ray spectrum from the central region, together with the integrated emission of all shells plus the central region is given in Figure 4, for the case where $E_{\max} = 10^8$ GeV.

## 5  Propagation to Earth

The universe is not transparent to $\gamma$-rays above about 100 TeV because of photon–photon pair production on the microwave background (see Protheroe 1986 and references therein). Below 100 TeV, interactions with the infrared background, due largely to infrared emission by nearby galaxies and redshifted optical emission from distant objects, becomes important for propagation over cosmological distances (Wdowczyk, Tkaczyk & Wolfendale 1972, Stecker *et al.* 1992, Protheroe and Stanev 1993). The interaction length in these fields is shown in Figure 5.

We consider the propagation of the $\gamma$-rays from the AGN to Earth taking account of a pair-Compton cascade in the infrared/microwave field as described by Protheroe and Stanev (1993). The spectrum corresponding to that given in Figure 4 for $E_{\max} = 10^8$ GeV is plotted in Figure 6 after propagation from AGN at several redshifts. The rejuvenation of the cascade by the inverse Compton process is quite noticeable for this relatively high maximum proton energy for redshifts less than $\sim 0.3$. However, inclusion of the inverse Compton process makes little difference to the spectra of sources which have significantly lower maximum proton energies (i.e. less luminous sources) and to the spectra of high-redshift sources.

For our results to be useful, we need to be able to scale them to AGN flux observations. For this we assume that the AGN continuum luminosity is a direct result of proton acceleration, and subsequent interactions; the electromagnetic component $(e^{\pm}, \gamma)$ resulting from proton–photon pair production and pion decay initiates an electromagnetic cascade responsible for producing the AGN continuum. We therefore divide the AGN continuum luminosity, $L_C$, by the energy per injected proton going into the electromagnetic component to obtain the rate at which protons are injected into the accelerator. The differential luminosity in high energy $\gamma$-rays from an individual AGN is then obtained from the total $\gamma$-ray emission per injected proton by multiplying by the proton injection rate.

We apply our result to predict the high-energy to very-high-energy $\gamma$-ray flux from the cores of three nearby AGN: NGC4151, Cen A and 3C273. For these nearby sources, this is simply a case of applying the inverse-square law together with a consideration of attenuation of $\gamma$-rays in the microwave and infrared background radiation fields. The integral flux expected from these individual AGN is given in Figure 7.

Above 100 GeV, all but the most optimistic of the flux predictions lie below the sensitivities of current Cerenkov and air shower arrays, and the model we use is therefore not in conflict with the observational data at the higher energies. The $\gamma$-ray fluxes above 100 MeV measured by EGRET prove a stricter test for our model, however, and are also given in Figure 7. For the blazar 3C273, there is no problem. The flux detected by Fichtel *et al.* (1994) lies above our prediction. This is not inconsistent with the present work as at least some, and possibly most, of the $\gamma$-ray emission is due to the jet. For the Seyfert galaxy NGC4151, the upper limit to the flux (Lin *et al.* 1993) is consistent with our model, but only if our most pessimistic flux is used. Whilst EGRET has not made a positive detection of the active galaxy Cen A (Schönfelder 1993), no published upper limit exists, and we assume here that it will be similar to that reported for NGC4151. This upper limit is some way below even our lowest flux prediction. While this is disturbing, one should bear in mind that our result is purely theoretical, and that because of the large uncertainties concerning the environment near AGN central engines, uncertainties of as much as a factor of ×10 are not unexpected for the γ-ray emission.

# 6   Integration over $L$ and $z$

If we know the local luminosity function of AGN, how the luminosity function evolves with redshift $z$, and the effective differential luminosity of γ-rays (taking account of cascading on the infrared/microwave background), $dL_\gamma/dE$ (eV s$^{-1}$ eV$^{-1}$), then we can perform an integration over redshift and luminosity to obtain the γ-ray intensity from AGN. For our assumed AGN continuum spectra, the 2 – 10 keV luminosity is related to $L_C$ by $L_X \simeq 0.05 L_C$. For the Einstein-de Sitter model ($q_0 = 0.5$) we obtain

$$\frac{dI_\gamma}{dE} = \frac{1}{4\pi}\frac{c}{H_0}E^{-1}\int_0^{z_{\max}} dz\, \frac{g(z)}{f(z)}(1+z)^{-\frac{5}{2}} \int dL_X\, \rho_0\left(\frac{L_X}{f(z)}\right)\frac{dL_\gamma}{dE}\{(1+z)E, L_X\}, \tag{26}$$

where $\rho_0(L_X)$ (cm$^{-3}$ (erg s$^{-1}$)$^{-1}$) is the local X–ray luminosity function of AGN, $f$ and $g$ describe the evolution of luminosity and number density in co-moving coordinate space, and we assume Hubble's constant to be $H_0 = 50$ km s$^{-1}$ Mpc$^{-1}$. We use the parameters of Model A of Morisawa *et al.* (1990) or the pure luminosity evolution model of Maccacaro *et al.* (1991) to describe the X-ray luminosity function of AGN and its evolution.

We perform our calculation for a range of $b$ from 1 to 100, for both spectrum A and spectrum B, and for both luminosity functions. In each case we assume that the population of AGN's has a distribution of shock radii such that $\log(x_1)$ is distributed uniformly between $x_1 = 10$ and 100, and integrate over $x_1$ accordingly. Our result is shown in Figure 8 for $b = 1$, 10, and 100, where the hatched bands give the range between maximum and minimum intensities calculated using spectrum A or spectrum B, and the Morisawa *et al.* or the Maccacaro *et al.* luminosity function.

The integration over redshift carried out above starts at $z = 0$ and we assume a uniform space density of AGN at a given redshift. This approximation becomes better with increasing redshift, and does not introduce any significant error into our calculation of the average intensity. However, it is possible that an anisotropic distribution of nearby AGN with fluxes just below detection threshold could give rise to an anisotropy in the measured $\gamma$-ray background intensity.

# 7 Conclusions

We would like to emphasize that our calculation is made using a simple model of AGN which may approximate the conditions in some AGN central engines in which accretion onto a supermassive black hole takes place. Our calculations do not apply to AGN models such as 'spinars' or 'magnetoids' although particle acceleration has been invoked in such models and would give rise to $\gamma$-ray production. Also, some AGN models do not permit proton acceleration up to the energies discussed in the present work, if at all. Clearly, the high energy $\gamma$-ray background calculated for such models would be much less than given in the present work which should therefore be considered as an estimate of the maximum contribution from AGN central engines, as distinct from AGN jets. That having been said, in Figure 8 we compare our result with the expected backgrounds at high energy due to interactions of cosmic rays in the microwave background and interstellar medium calculated by Halzen *et al.* (1990) and Berezinsky *et al.* (1993) respectively. Recently, Padovani *et al.* (1993) and Stecker *et al.* (1993) have estimated the contribution of blazars to the diffuse $\gamma$-ray background. The calculation of Stecker *et al.* was based on EGRET observations of AGN, together with a correlation between γ-ray and radio luminosities. However, their published estimate is incorrect because they had assumed the radio luminosity function they used (Dunlop and Peacock 1990) gave densities per proper volume, whereas it gave densities per co-moving volume. This has been corrected by one of us (personal communication from R.J.P. to F.W. Stecker) and is shown in Figure 8 along with the estimate of Padovani *et al.* (1993).

We have added to Figure 8 the SAS-2 γ-ray background observations (Thompson and Fichtel 1982) and note that at 100 MeV our results are a factor of ×2 to ×8 above the data. While the blazar calculations are based on observed γ-ray fluxes of AGN, our result is purely theoretical and, for reasons given earlier, uncertainties of as much as a factor of ×10 are not unexpected for the γ-ray emission. Such uncertainties would apply less to the neutrino and cosmic ray output which is not so sensitive to the AGN environment.

The estimated diffuse γ-ray backgrounds due to both blazars and central engines of 'ordinary' AGN (present work) are comparable at 10 GeV, with the central engine contribution decreasing more rapidly with increasing energy. The high blazar γ-ray intensity predicted by Stecker *et al.* at TeV energies results from their assumption that all blazars produce an $E^{-2}$ spectrum extending beyond 1 TeV.

Finally, we conclude there is likely to be an important contribution to the diffuse $\gamma$-ray background from 'ordinary' AGN as well as blazars, cosmic ray interactions in the interstellar medium, and cascading from ultra-high energy cosmic ray interactions with the cosmic microwave background radiation.

Figure 1: The maximum proton energy, $E_{\max}$, vs. the AGN continuum luminosity, $L_C$, divided by $b^2$ where $b$ is the ratio of the diffusion coefficient to the Bohm diffusion coefficient. The relationship is given for both types of continuum: spectrum A (heavy shading) and spectrum B (light shading). The upper and lower limits in each case correspond to $x_1 = 100$ and $x_1 = 10$ respectively.

Figure 2: The spectra of $\gamma$-rays (dashed lines) and electrons (full lines) produced per injected proton are given for various spherical shells. Results are shown for $E_{\max} = 10^8$ GeV, Spectrum A, $x_1=30$ and $b = 100$. The numbers attached to the curves give $\log(r_{\rm in}/R)$, where $r_{\rm in}$ is the inner radius. In each case the outer radius is $r_{\rm out} = 10^{0.5} r_{\rm in}$. The thick curves show spectra of $\gamma$-rays and electrons produced in the central region.

Figure 3: The average escape probability $P_{\rm esc}$ is given for various radii for $\gamma$-rays in the spectrum A photon field for $x_1=30$ [solid curves labelled by $\log(r/R)$]. The average escape probability for a $\gamma$-ray produced inside the central region (dashed line) is also given.

Figure 4: The $\gamma$-ray emission per injected proton from various shells and the central region is given for the case where $E_{\max} = 10^8 \text{GeV}$, $x_1=30$, $b=100$ and spectrum A was used. The dotted line indicates the integrated emission from all shells and the central region. The dashed lines indicate the expected emission without the contribution of the pair–synchrotron cascade. The labels attached to the curves have the same meaning as in Figure 2, and again the thick lines indicate the contribution from the central region.

Figure 5: The interaction length of $\gamma$-rays in the infrared and microwave background due to pair-production (from Protheroe and Stanev 1993). We take the temperature of the microwave background to be 2.735 K, and for the infra-red bacground we use the spectrum given in Fig. 1 of Protheroe and Stanev (1993).

Figure 6: The spectrum of $\gamma$-rays per injected proton after cascading in the microwave background from an AGN with an $E_{\max} = 10^8\,\mathrm{GeV}$, spectrum A and $b=100$ (thin solid lines). The curves are (from top to bottom) for redshifts of: 0.001, 0.003, 0.01, 0.03, 0.1, 0.3, 1.0, 2.0, 3.0, 4.0 and 5.0. The equivalent curves are given as dashed lines when only absorption is taken into account, and for comparison, the spectrum without any attenuation is also given (thick solid line). (We assume $H_0 = 50$ km s$^{-1}$ Mpc$^{-1}$.)

Figure 7: The expected integral flux of $\gamma$-rays after attenuation in the infrared and microwave background from three separate AGN: NGC 4151 (horizontal hatching), 3C273 (vertical hatching) and Cen A (diagonal hatching). EGRET detections and upper limits are also given, a triangle, square and diamond indicating 3C273, NGC4151 and Cen A respectively. The predicted and measured flux from 3C273 has been scaled up, and that for Cen A down, both by a factor of $10^4$, to improve clarity. The spread in the predicted flux in each case arises from varying the unknown parameters, $b = 1 \rightarrow 100$, $x_1 = 10 \rightarrow 100$, and by using both spectrum A and B.

Figure 8: The expected diffuse $\gamma$-ray flux for $b=1$ (horizontal hatching), $b=10$ (thin oblique hatching) and $b=100$ (thick oblique hatching). In each case, the width of the band shows the extreme range of intensity between results obtained using the two luminosity functions and using the two AGN continuum spectra. An integration over a flat distribution in $\log x_1$ has been made for $10 < x_1 < 100$. The contribution expected from blazars estimated by Padovani *et al.* (1993) (solid line), and based on the work of Stecker *et al.* (1993) (dotted line; see text) is shown. The expected contributions from interactions of cosmic rays in the microwave background (Halzen *et al.* 1990) ($p\gamma$) and in the interstellar medium (Berezinsky *et al.* 1993) ($pp$) are shown. The vertical hatched band at low energies shows the intensity observed by SAS-2 (Thompson and Fichtel 1982).

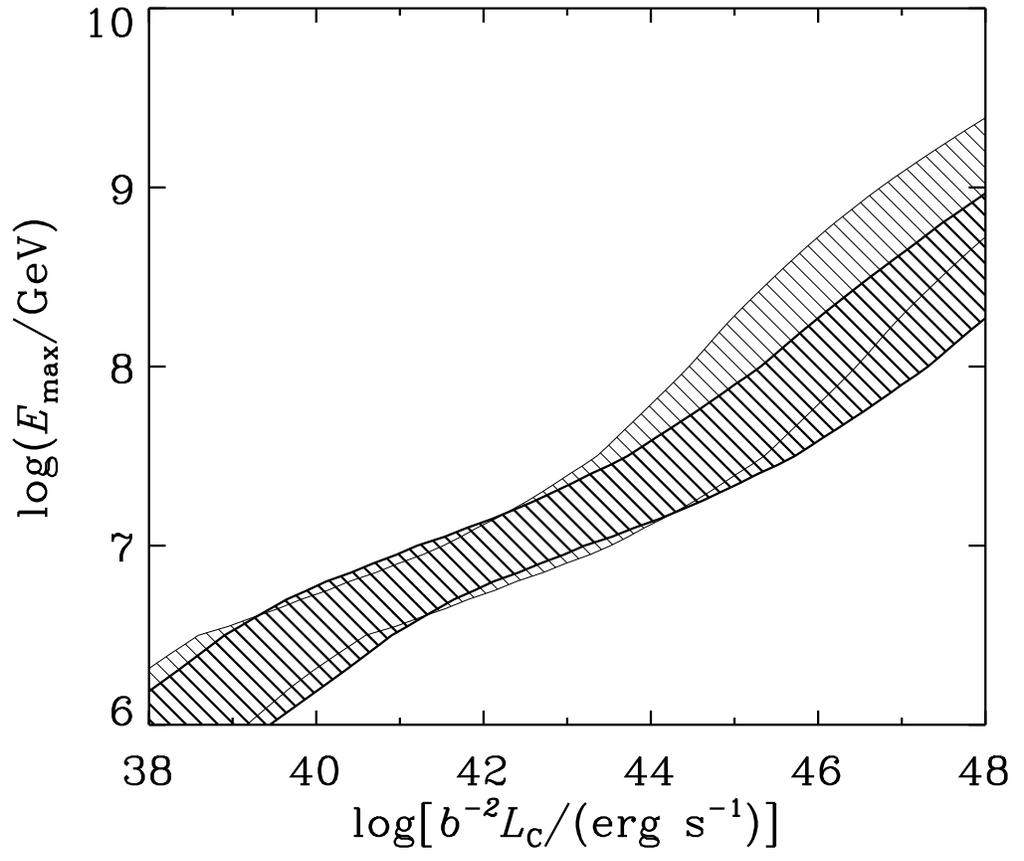

Figure 1

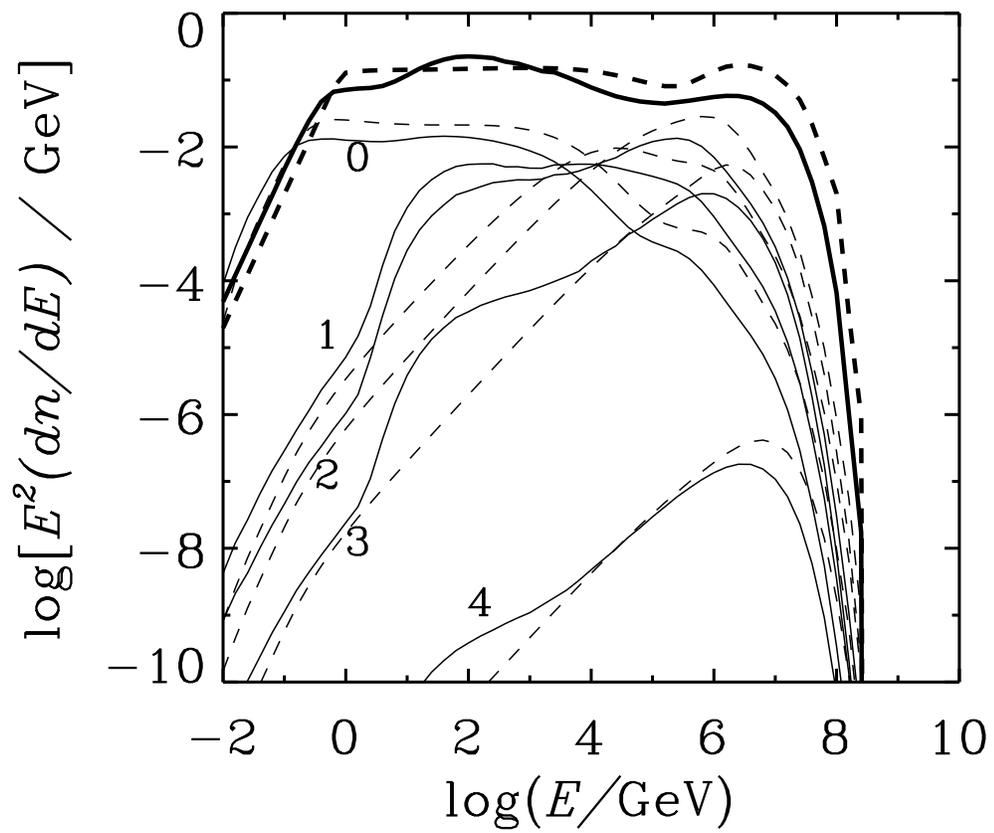

Figure 2

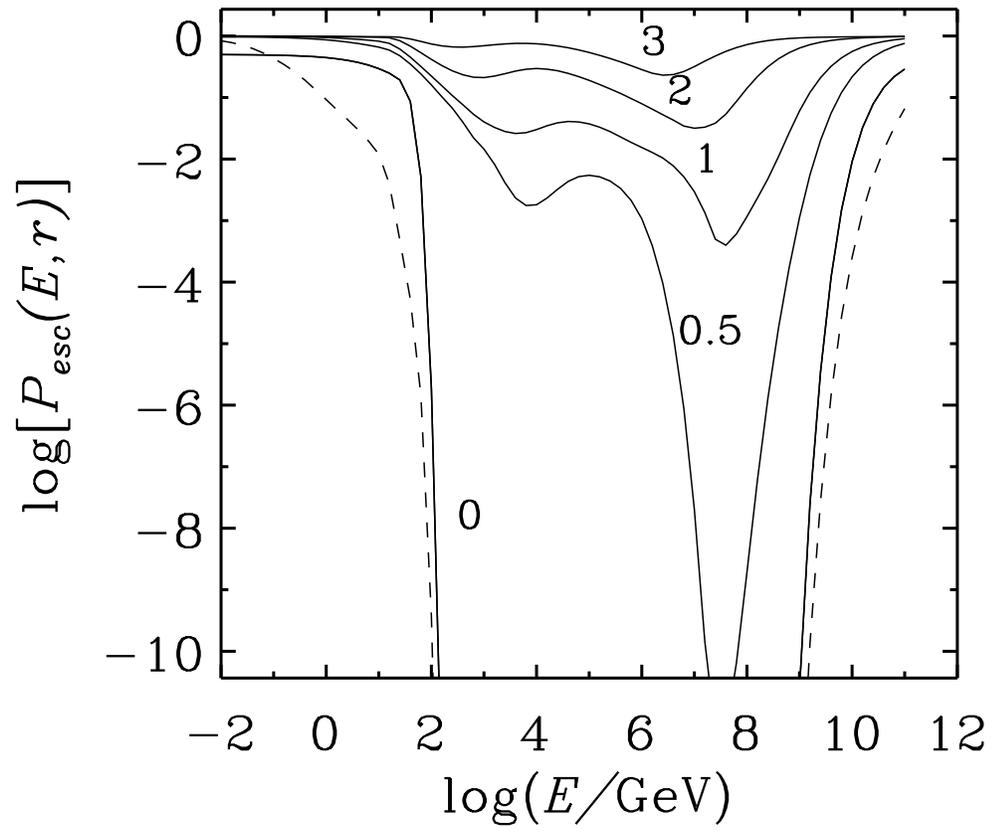

Figure 3

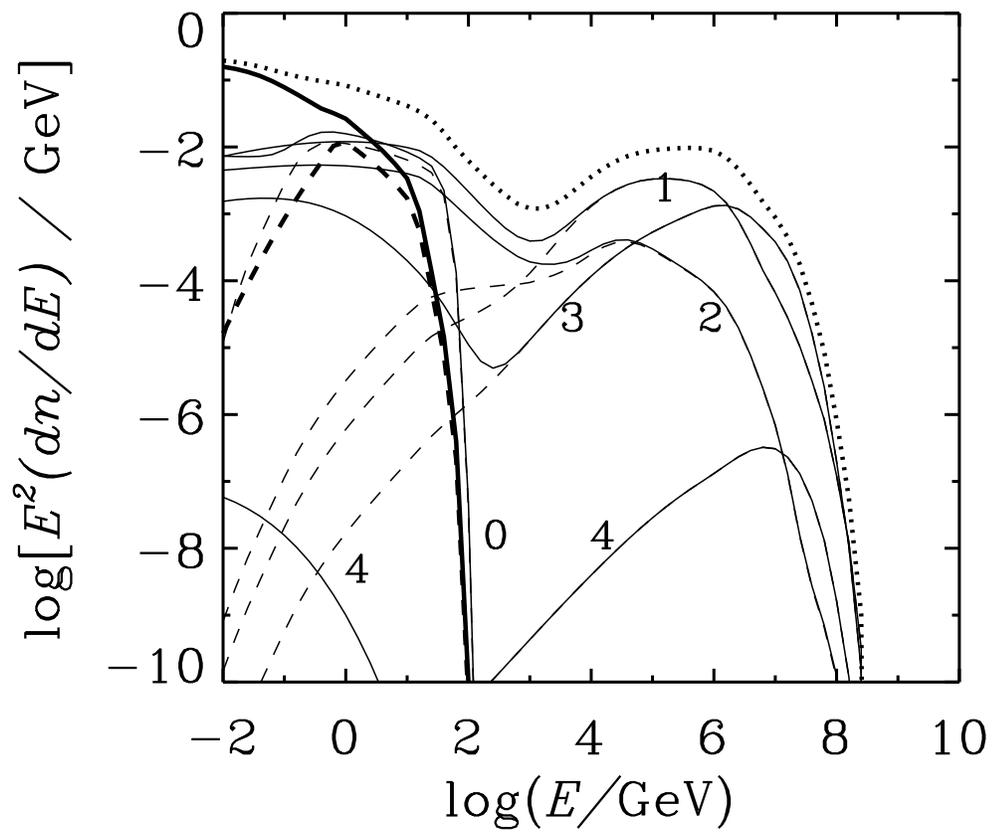

Figure 4

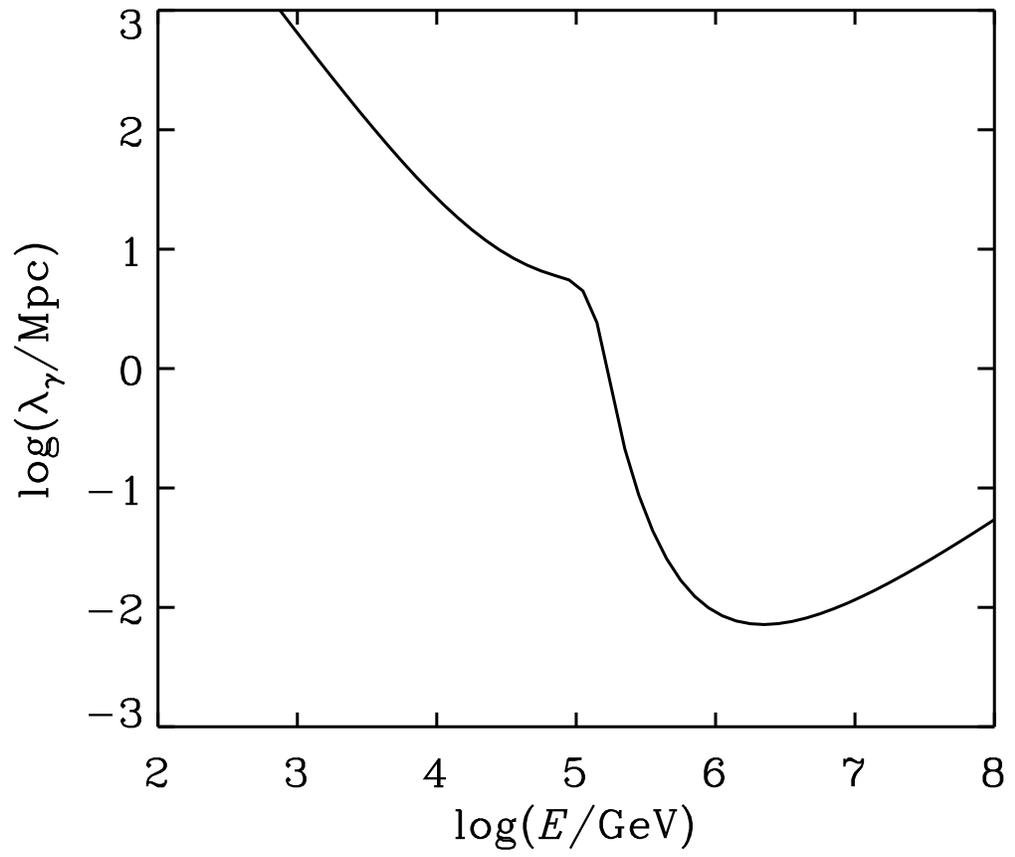

Figure 5

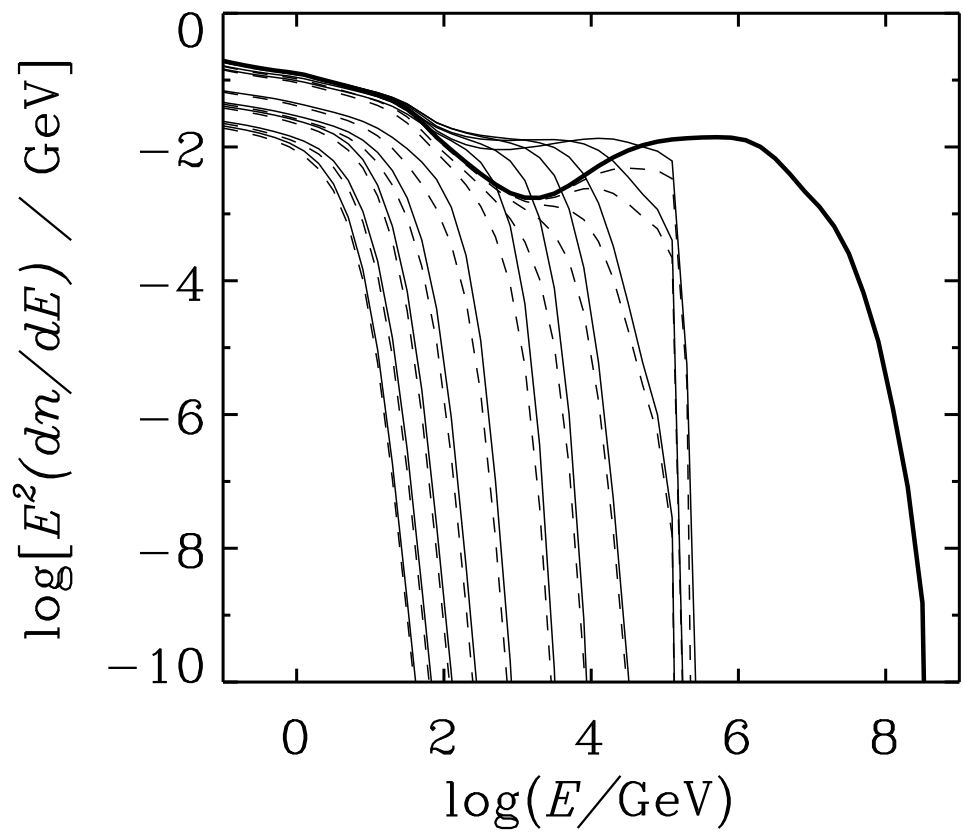

Figure 6

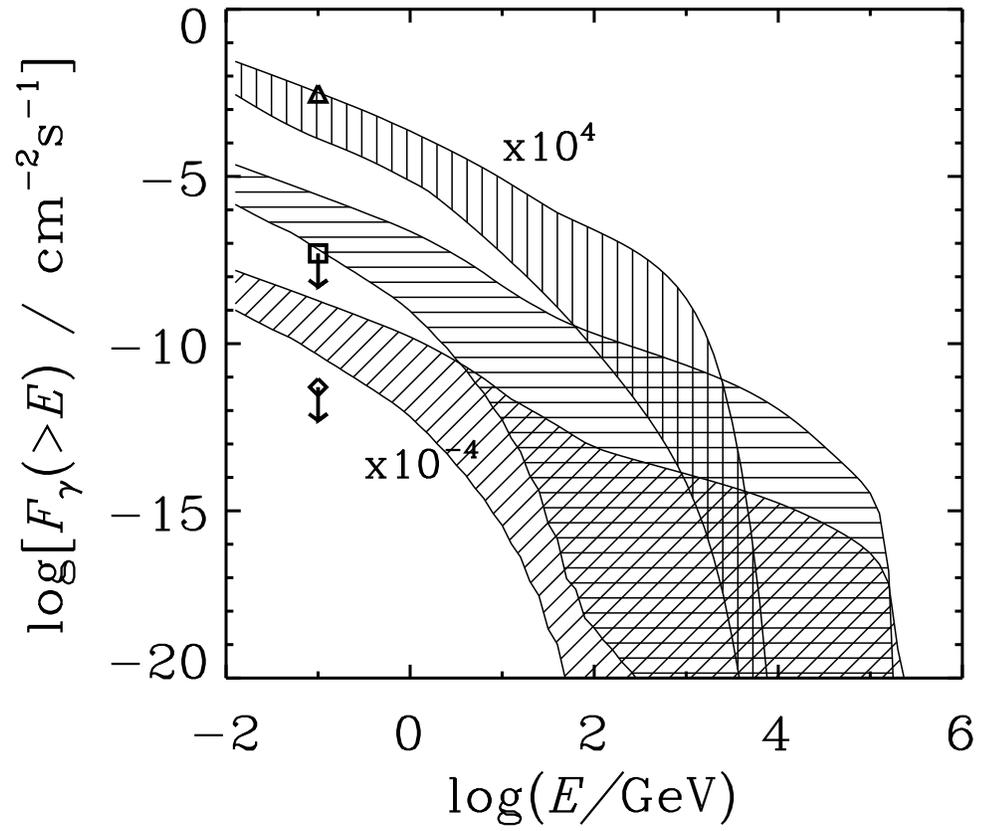

Figure 7

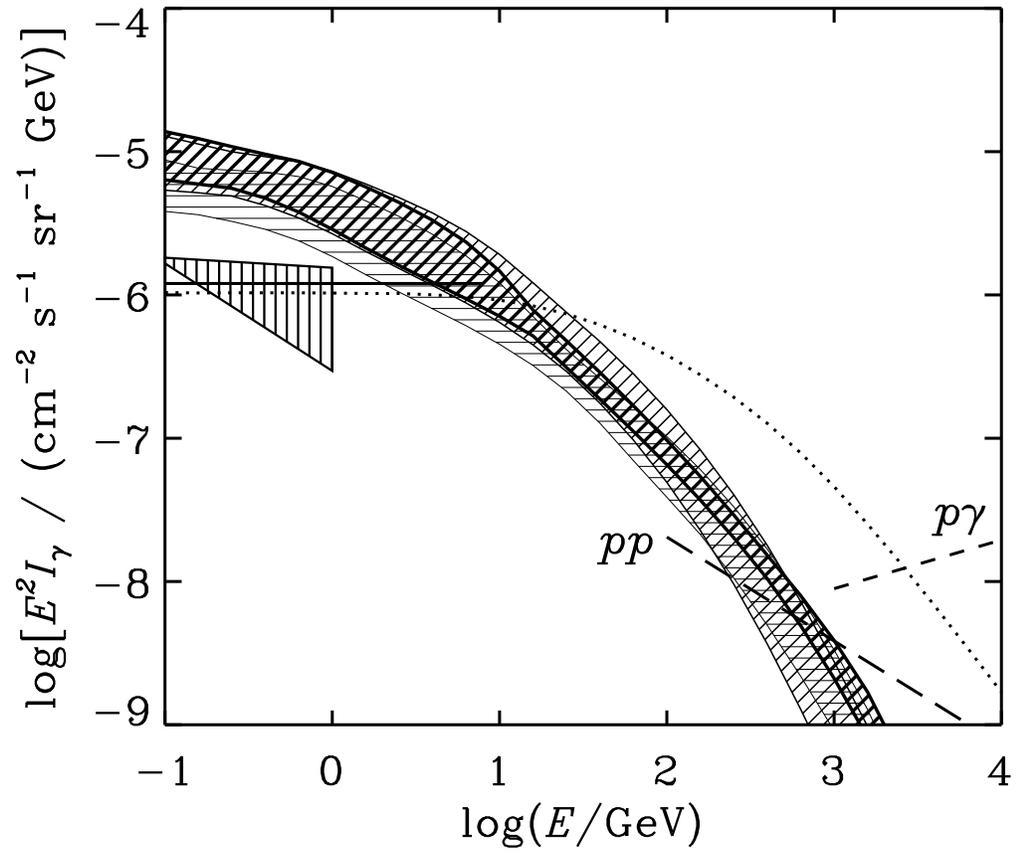

Figure 8